\documentclass[12pt,preprint]{aastex}
\usepackage{graphicx}
\usepackage{amssymb}
\newcommand{\swift}{SWIFT J1753.5$-$0127}
\newcommand{\xte}{XTE J1118+480}
\newcommand{\gx}{GX~339$-$4}

\begin{document}

\title{\swift: a surprising  optical/X-ray cross-correlation function}
\author{Martin Durant}
\affil{Instituto de Astrof\'isica de Canarias, La Laguna, E38205 Tenerife,
  Spain}
\email{durant@iac.es}
\author{Poshak Gandhi}
\affil{ RIKEN Institute of Physical and Chemical Research, 2-1 Hirosawa,
  Wakoshi, Saitama, Japan}
\author{Tariq Shahbaz}
\affil{Instituto de Astrof\'isica de Canarias, La Laguna, E38205 Tenerife,
  Spain}
\author{Andy Fabian}
\affil{Institute of Astronomy, Madingley Road, Cambridge, UK}
\author{Jon Miller}
\affil{University of Michigan, 500 Church Street, Ann Arbor MI, USA}
\author{V. S. Dhillon}
\affil{Department of Physics and Astronomy, University of Sheffield,
  Sheffield S3 7RH, UK}
\and
\author{Tom R. Marsh}
\affil{Department of Physics, University of Warwick, Gibbet Hill Road,
  Coventry CV4 7AL, UK}

\begin{abstract}
We have conducted optical and X-ray simultaneous observations of
\swift\ with RXTE and ULTRACAM, while the system persisted in its
relatively bright low/hard state. In the cross-correlation function (CCF),
we find that the optical leads the X-rays by a few seconds with a
broad negative peak, and has a smaller positive peak at positive
lags. This is markedly different from
 what was seen for the similarly interesting system \xte, and the
first time such a correlation function has been so clearly
measured. We suggest a physical scenario for its origin. 
\end{abstract}
\maketitle

\section{Introduction}
X-ray and optical emission from astrophysical objects are produced by
very different means, with different energetics and time-scales. X-ray
binaries are prolific sources of both, with measurable rapid
variability. The emission modes cannot be wholly independent, so by
comparing their inter-connection, we can learn about the physical
conditions from the main emission region: the inner disc around a
neutron star or black hole.

Whilst high time resolution ($\leq1$\,ms) for X-ray observations has
been achieved from the very earliest observations, the photon rates
for most sources prevented statistically significant timing work. In
the optical domain, exposure times in the ms domain have only become
possible relatively recently, with low enough noise and dead-time and
high enough quantum efficiency (i.e., CCDs as opposed to photometers)
to achieve good signal-to-noise ratios per exposure. The final problem
has been simply to schedule simultaneous X-ray and optical
observations. 

The black hole X-ray Transient (XRT)  \swift\ is a system which has been of
great interest recently following its outburst episode and
detailed observations with the {\em SWIFT} satellite. First discovered
by the {\em SWIFT}/BAT (Burst Alert Telescope; Palmer et al., 2005)
in 2005, pointed $\gamma$-ray, X-ray, UV, 
optical and radio observations all detected a new bright source at
this location (Morris et al., 2005; Still et al., 2005; Halpern  2005;
Fender et al., 2005). Following the early report of a 
0.6\,Hz quasi-periodic oscillation (QPO; Morgan et al. 2005; 
Ramadevi \& Seetha, 2005), persistent for some time
after the bursting episode, we applied to observe the system
simultaneously in X-rays and optical as it faded\footnote{ESO and RXTE
observation IDs 079.D-0535 and 93119-02-02-00, respectively.}. The
source was still relatively bright however, especially in the optical,
at the time of our project. Following the burst, Cadolle-Bel et
al. (2007) measured the spectrum from radio up to 600\,keV, and Miller
et al. (2006) showed spectroscopically that a disc reaching down to
small radii was likely. 

Further details of our observational campaign, including spectroscopy,
longer-term
multi-band optical photometry and detailed periodogram analysis are to
be published separately in Durant et al. (2008), and an
analysis of the long-term R-band variability and 
orbit-like modulation are presented in Zurita et al (2008). Here, we
intend to make the minimum number of processing steps 
and assumptions to obtain the X-ray/optical CCFs
a unique phenomenological hint of black hole accretion
physics.  

\section{Observations}
\swift\ was observed for 53.6\,min on 13 Jun
2007 with the {\em Rossi X-ray Timing Explorer} (RXTE), which provides
very high timing resolution ($\sim1\,\mu$s), reasonable energy
resolution ($\sim1$\,keV) and very high effective area in the
2--100\,keV range. During this
observation, three of the units of the Proportional Counting Array
(PCA, Bradt et al. 1993) were active. We do not consider here the data
of the HEXTE or ASM instruments, where the count rates were much
lower. The 2--20\,keV flux was
1.6$\times10^{-9}$\,erg\,cm$^{-2}$s$^{-1}$ (with standard background
subtraction). 

We produced four light-curves of the data using the FTOOLS task {\em
  seextrct} with standard good time filtering and standard
settings. The bands were selected purely by splitting the energy
channels into four equal segments. We have not attempted to calibrate
these energy cuts exactly: the rough general mapping of channel number
to energy at {\tt heasarc.nasa.gov/docs/xte/e-c\_table.html} is
sufficient for our uses. The variability was at the
$\sim40$\% level (in which Poissonian noise is significant).

\swift\ was simultaneously observed with ULTRACAM,
mounted on VLT/3 (Melipal) telescope. Of the 75\,min observation,
$\sim$50\,min were 
simultaneous with the RXTE observation. ULTRACAM is an instrument
employing dichromatic beam  
splitters, frame-transfer CCDs and a GPS-based timing system in order
to be able to make simultaneous multi-wavelength light-curves at very
high time resolution, up to 300\,Hz (Dhillon \& Marsh, 1999; Dhillon,
2007). We used 
two small windows on each CCD (one for the 
source of interest, one for a local standard), with exposure times of
140\,ms (resulting in a duty cycle of 142\,ms). Here we note that
conditions were fairly poor, with thin 
cloud causing transparency and seeing variations, mostly on timescales
longer than 10\,min. 
Fluxes were extracted by aperture photometry with a variable aperture
size scaled to the FWHM of the reference star on each image. This
enables some optimization for signal to noise under the variable
conditions. Short-term variability in the optical band was at the
$\sim$10\% level.

\section{Cross-correlation}
\begin{figure}
\begin{center}
\includegraphics[width=0.48\hsize]{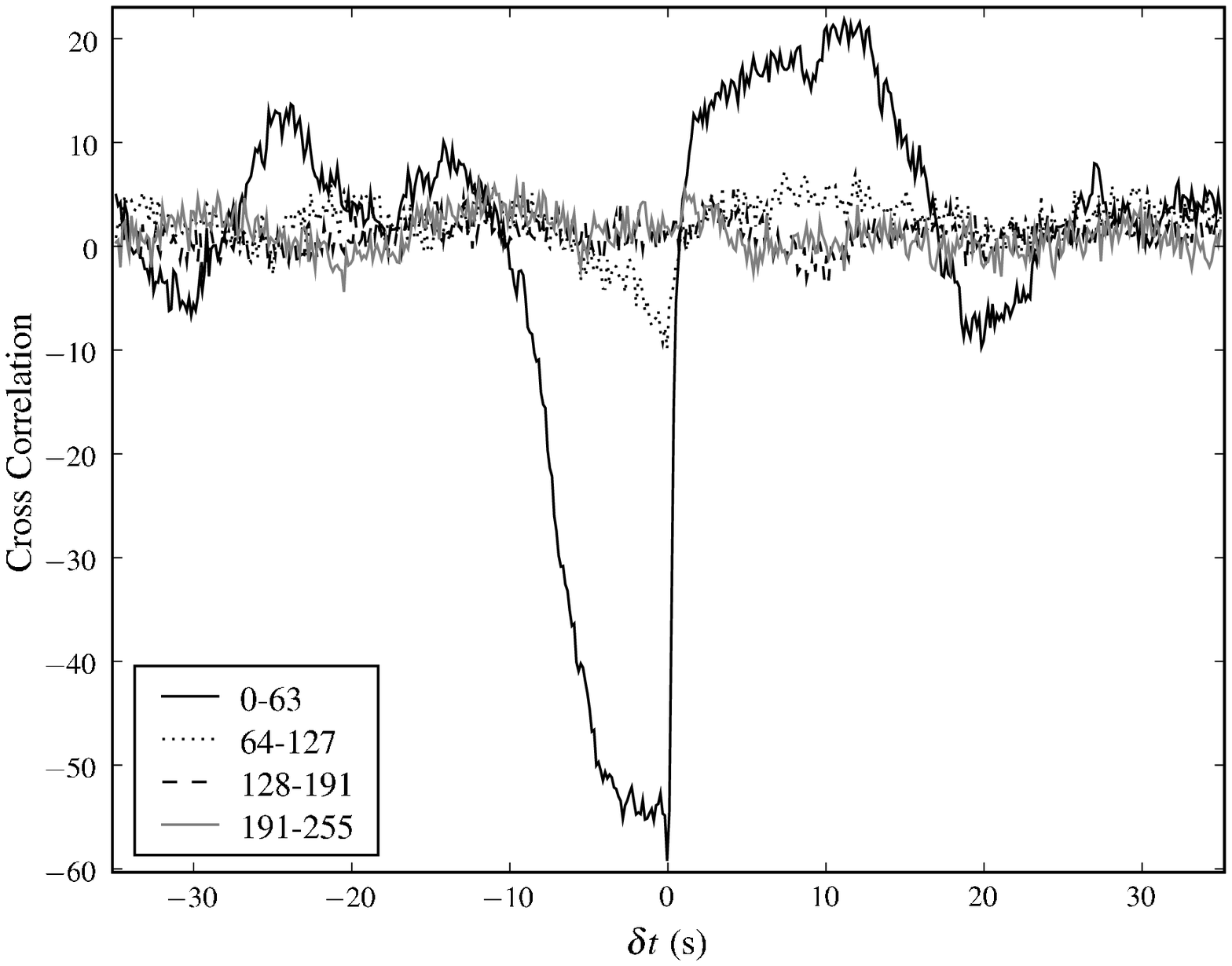}
\includegraphics[width=0.48\hsize]{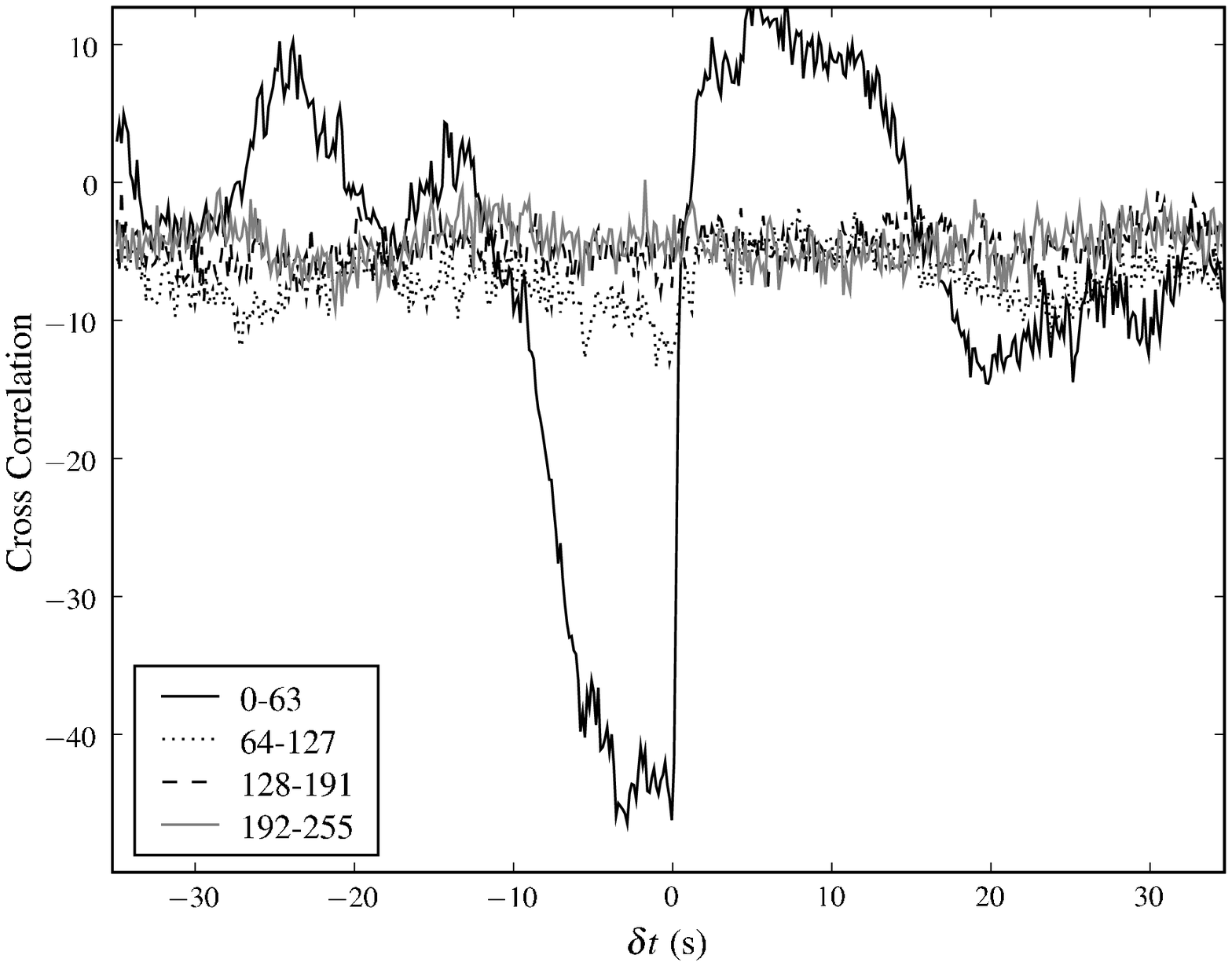}
\caption{Cross correlation of RXTE/PCA and ULTRACAM light-curves of
  \swift. $\delta t<0$ represents optical light arriving earlier than
  X-rays. Each plot shows four X-ray energy ranges, by PCA channel 
  number, for the r' band (left) and for the g' band (right). Scaling
  is relative to the noise in each CCF (see
  text). }\label{cross}
\end{center}
\end{figure}

We calculated the CCFs of each of the X-ray
light-curves with each optical light-curve produced above.\footnote{The
light-curves themselves, periodograms and auto-correlations will be
published separately (Durant et al., 2008).} The results
can be seen in Figure \ref{cross}. The curves have been scaled such
that one unit equals the typical difference between one point and the
next - it is therefore a simple measure of the random ('white')
noise, as seen in the cross-correlation. The scale is therefore a
simple measure of significance.

Whilst the functions are clearly
very similar between optical bands (except that the g' band is much
noisier), there is a marked difference between the behavior seen with
X-ray energy range:-
\begin{itemize}
\item{For the 0--63 channel range, which in Epoch 5 corresponds
  roughly to 2--27\,keV, a strong signal is
seen at negative lags (optical leading X-rays), followed by a somewhat
weaker inverted signal. The width of these signals is of the order
10\,s. There is further, less significant structure in the CCF apart
from the two main features mentioned. This could hint at some
oscillatory interaction between the optical and X-ray emission.}
\item{In the 64--127 channel range ($\sim$27--55\,keV), a weaker but
  significant signal is  seen as for 0--63, but notably narrower in
  its response.} 
\item{For the two remaining curves at higher energies
  ($\sim$55--118\,keV), no significant signal is seen at all.}
\end{itemize}
In every CCF, no further significant features are seen for 
$|\delta t|>30$\,s.
The g' band may show a broader response than the r' band. With the
poorer data in the former, we regard this as merely a suggestion. 

Significantly, the cross-correlations here do not change noticeably
during the simultaneous window, when constructed for sections of the
data. This is despite the window length ($\sim$50\,min) being
significant compared to the orbital-like modulation period of 3.2\,h
(Zurita et al. 2008).

To investigate what features of the light curves are responsibe for
the CCFs in Figure \ref{cross}, we performed the following check. For
each of the 1\% brightest bins of the X-ray total light curve, we
averaged  windows of the optical light curve centred on these
bins. The resulting average looks like the solid line in
Figure \ref{cross} (the X-ray light curve is, of course, dominated in
terms of counts by the lower energy range). Conversely, for the
average of windows of the optical light curve centred on the 1\% {\em
  faintest} X-ray bins, the average function is again like the solid
curve, but with the y-axis flipped. In other words, the optical light
curve, some seconds before the X-ray, tends to be going in the
opposite direction; yet the optical a few seconds after the X-ray
tends to be going in the same direction, somewhat weaker. Since
neither light curve can be decomposed into discreet events or flares,
this relationship only comes out over an average of the whole light
curve. 

\begin{figure}
\begin{center}
\includegraphics[width=0.48\hsize]{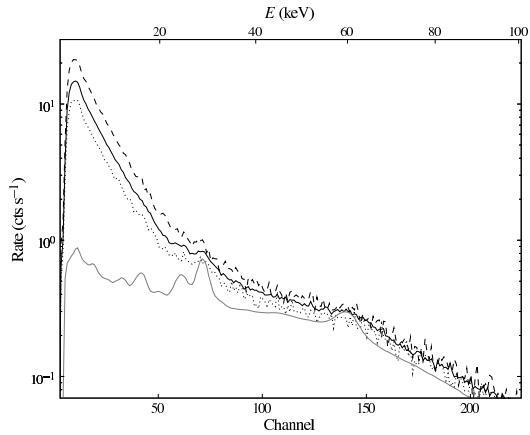}
\caption{Counts spectrum extracted for the RXTE/PCA observation of
  \swift, uncalibrated: mean spectrum (solid line), high count rate
  bins (dashed line) and low count rate bins (dotted line). Note the
  logarithmic vertical scale. For comparison, we also plot the
  estimated background spectrum (gray), see text. Note how the
  background curve approaches but does not exceed the average
  spectrum. }\label{spec}
\end{center}
\end{figure}

We constructed minimum-assumption spectra from the data, to
investigate further which component is varying. Figure \ref{spec}
shows the count rate per energy channel for the $\sim$10\% brightest
and faintest bins, along with the spectrum for all bins. One sees that
from the lowest energies measured ($\sim$2\,keV) to about Channel 70
($\sim$30\,keV), the
spectrum maintains the same form, fluctuating by 
about a factor of 2. Up to Channel 130 ($\sim$60\,keV), some variation is seen,
but none at higher energies. Thus it is not surprising that the higher
energy bands above do not correlate at all with the optical. Note that
the count-rates are clearly much lower in this upper energy range, and
furthermore that the background rate becomes dominant. It is not
surprising, therefore, that Figure \ref{cross} shows no relationship
in comparison to the white noise above $\sim$60\,keV. There are,
however, still some counts above the background for this hard
source. We compared the ratio of the sum of background-subtracted
counts in the High and Low curves above and below channel 127, and
find that they are very similar, perhaps the high-energy part showing
an even higher ratio (i.e., it varies more). In general, it seems that
the spectrum varies in amplitude only and not in shape.
The modelling of background spectra has been a rapidly changing
calibration issue\footnote{\tt
  http://www.universe.nasa.gov/xrays/programs/rxte/pca/doc/bkg/bkg-2007-saa/},
however, so we regard this as suggestive. We used the latest
calibration data available in April/May 2008. 

\section{Discussion}
Very few X-ray/optical CCFs are recorded in the literature, mostly
of systems in low states, where the optical lags the X-rays, and
is assumed to be the reprocessing signature, possibly from the large
inner radius of a truncated accretion disc. This can be used for
tomography of the disc and companion by tracking the lag evolution
with spectral range and orbital phase (see particularly Hynes, 2005).
The dearth of cross-correlations is in good part due
to the logistics of arranging simultaneous observations with the few
instruments capable, within the short window following an outburst. In
this respect, \swift\ has been unique, by persisting in its low/hard
state (relatively bright) for years after outburst (Zurita et
al. 2008). We may find that 
similar relations exist in other systems, which have gone undetected
for technical reasons. Notably, Hynes et al. (2006)
report the optical lagging the X-rays in this system by $\delta
t<10$\,s during outburst, so clearly it was in some different mode
during our observations.

Most typically, one expects X-ray flaring to occur in the innermost,
hottest regions, and power optical emission by reprocessing
of the X-ray flux further out in the optically thick accretion
disc, or on the surface of the companion. This picture is a natural
consequence of the typical ADAF model geometry, where the disc
truncates at a large 
inner radius, inside of which material is hot and low-density (the
ADAF itself), where X-rays are produced. One would expect from this a
CCF positive 
response with a steep rise and slower fall. The positive part of our
CCFs do not show this, so simple reprocessing does not dominate over
the positive lag region. 

Alternatively, X-ray emission may be from a magnetically driven corona
around the dense accretion disc (ADC), where reconnection produces
energetic particles, and the energy release is dominated by
hard X-rays, from the up-scattering of photons by these highly
energetic coronal particles. The different energy outputs can thus be
coupled and interrelated in a complex manner, e.g., the jet-disc
coupling model of Malzac et al. (2004). The corona can form a self-limiting
feedback system (see e.g., Uzdensky \& Goodman 2007, 2008)
 wherein particles can evaporate from the disc to the
corona or condense back to the disc, and the reconnection rate is
determined by the density of the corona and magnetic loop
movement. Reconnection occurs preferentially in a {\em marginally
  collision-less} coronal medium where free magnetic and particle
kinetic energies are comparable, which results in a stable equilibrium for
a given parameter set (energy transfer rate, sheering etc.).

If the emission is composed of localized micro-flares ({\em flickering}), each
flare might proceed thus: between flares, a patch of the accretion
disc cools, and particles from the corona are able to condense; as the
coronal density drops, it becomes less collisional and magnetic energy
dominated towards the marginally collisionless state, where particle
density cannot inhibit magnetic reconnection. At a critical point,
cool X-rays flash at the moment of 
reconnection, decreasing the stored magnetic energy; the disc is
heated again, and the coronal particle 
population replenished by evaporation. Thus, the X-ray emission of the
micro-flare is timed after a period of condensation, when the optical
emission would be decreasing as more particles are shielded behind 
cyclotron absorption (indeed, free particles in the corona may also
radiate by cyclotron/curvature); immediately afterwards, a population of corona
particles and possibly the surface area of the disc patch is larger,
so the optical emission is enhanced again. Only a fraction of the released
X-rays and accelerated particles heat the disc, most escape or are
emitted at higher energies. The
self-regulating aspect,, in which the X-ray luminosity acts in the
opposite sense and reverses a dip in the optical; magnetic energy
driving and disc 
evaporation/condensation do seem consistent with our CCF. How magnetic
field is transmitted from the disc to the corona is not known.

This agrees roughly with the model by Fabian et al. (1982), where
emission is controlled by the growing and contraction of optically
thick, cyclotron-emitting plasma clouds.  This model fitted well with
the early measurements of \gx\ (see below). Since the energy emitted
by cyclotron depends on the electron number and magnetic energy
density, as field is expelled from dense regions, the optical emission
would decrease while the energy available for reconnection
increases. If the cyclotron emission is
predominantly optically thick, then the emitted luminosity in the
optical depends on the surface area of these clouds - as they 
contract and expel magnetic field to the more tenuous, hot medium, the
optical emission decreases as the X-ray emission increases, they are
naturally anti-correlated. 

If the optical emission is not, then, predominantly jet-like, then 
there is no longer a reason to expect radio emission to correlate with
luminosity: Cadolle-Bel et al. (2007) noted (confirmed by Soleri et
al. 2007) that the radio emission was unusually low, and that its
synchrotron-like spectrum fell below the optical emission. Furthermore,
reprocessing of the emitted X-rays will be affected by the
acceleration of plasma in magnetic reconnection events. If there is
mildly relativistic bulk motion away from the denser matter, then the
reprocessing would be weakened and show a different time-response
(Beloborodov, 1999).

In the recent work by Liu et al (2007), they show that \swift\ in
particular, and also \gx\ can be modelled as a cool inner disc where
thermal conduction and Compton cooling are important in this disc's
interaction (condensation and evaporation) with the surrounding
low-density corona. They do not specifically consider timescales and
driving in their model, but their model is at least consistent with
the results here and similar to the picture presented above, and to
the dynamic picture of Fabian et al. It will be interesting to see
further development of their model. 

In our source, the high-energy  emission dominates the
total luminosity (up to INTEGRAL energies, see e.g., Cadolle-Bel et
al. 2007). The picture is, therefore, of emerging magnetic flux from
the disc, releasing its energy in the corona in a self-regulating
way. Optical emission is from the dense, hot, magnetically active disc
(by cyclotron/synchrotron) and by particles in the corona (by
cyclotron and
curvature). We believe this scenario accounts qualitatively for what
is seen, but is short of a proof.

\subsection{Comparison with \xte\ and \gx}
Two objects that have been extensively studied, including simultaneous
X-ray/optical projects, are \xte\ and \gx.

\xte\ seems initially very similar in many characteristics to \swift:
persistent low/hard state, high galactic latitude, X-ray spectral and
timing characteristics. Kanbach et al. (2001) found that the optical
emission lags the X-rays by a small amount ($\sim$0.5\,s), but there is an
interesting 'precognition dip' in the CCF which is
difficult to explain. These features may be qualitiatively similar to
our CCF, but in our case,
the main feature is a strong anti-correlation, optical before X-ray. The
alternative view would be that we see the same precognition dip
and response signal, but with very different
intensities. Interestingly, they find that the dip is stronger for
longer wavelength optical data, which would also fit our data (with
the caveat on the quality of the g' band observations above). It was
to describe this system that Malzac et al. (2004) developed their jet-disc
coupling model.

In the earliest such measurement made that we are aware of, Motch et
al (1983) derived the X-ray/optical CCF for
\gx. From a very short simultaneous observation window (96\,s), they
suggested a optical-leading anti-correlation, but only at energies
$E<13$\,keV. It was in this context that the model of Fabian et
al. (1982, above) was fairly successful. The CCF was not independently
confirmed, but appears similar 
to our work. Later, Gandhi et al. (2008 in prep.) repeated these measurements
over a longer time base-line for the source presumably in quiescence,
and found that the CCF similar to \xte, with the strongest feature a
weak positive peak 
showing a slight lag of optical behind X-rays (by $\sim$0.2\,s), but
the peak has a markedly different shape with a shallow rise and steep
fall, followed by negative correlation in the 1--3\,s lag range.

These comparisons are suggestive that the CCF we find is symptomatic
of the accretion mode in our object, at the time of observation.

\subsection{Conclusions}
Notwithstanding the technical difficulties of constructing
X-ray/optical cross-correlations, of the few capable instruments and
simultaneous scheduling, this work presents the functions for
\swift\ which challenge our understanding of the physical processes in
the immediate vicinity of a black hole. We find a strong
anti-correlation, with the optical preceding the X-rays on
time-scales of 1-10\,s. This demonstrates that there
exists a causal link between the optical and X-rays, aside from simple
reprocessing, and detailed dynamical modelling will be required to
describe the system more fully.

\medskip\noindent{\bf Acknowledgements:}
MD and TS are funded by the Spanish Ministry of Science.
PG is a Fellow of the Japan Society for the Promotion of Science
(JSPS). ULTRACAM was designed and built with funding from PPARC (now
STFC), and used as a visiting instrument at ESO Paranal, and RXTE is
operated by NASA. 
Partially funded by the Spanish MEC under the
Consolider-Ingenio 2010  Program grant  CSD2006-00070: ``First Science
with the GTC''  ({\tt http://www.iac.es/consolider-ingenio-gtc/}).

\end{document}